\documentclass[letterpaper, 10 pt, conference]{ieeeconf}  

\IEEEoverridecommandlockouts                              

\usepackage{graphicx} 
\usepackage{amsmath} 
\usepackage{cite}
\usepackage[utf8]{inputenc}

\usepackage{xurl}
\usepackage[hidelinks]{hyperref} 

\title{\LARGE \bf
Industry case studies of adapting TTS for smooth interactions
}

\author{Paige Tuttösí$^{1,2}$, El\'{e}onore Ferrier-Barbut$^{1}$, Anais Scipioni$^{1}$, Olivia Yung$^{3}$, Ha Eun Shim$^{3}$, Ajla Basic$^{4}$, \\ Yue Wang$^{3}$, and H. Henny Yeung$^{3,5}$
\thanks{ This work was supported by the Simon Fraser University FASS Breaking Barriers Interdisciplinary Incentive Grant, the France Canada Research Fund and the Rajan Family}
\thanks{$^{1}$P.Tuttösí, E. Ferrier-Barbut, and A.Scipioni are with Enchanted Tools, 18 rue de la Fontaine au roi, Paris, France, {\tt\small paige@enchanted.tools}}%
\thanks{$^{1}$P.Tuttösí is with the School of Computing Science,
        Simon Fraser University, 8888 University Drive, Burnaby, BC, Canada}%
\thanks{$^{3}$O.Yung, H.E.Shim, Y.Wang, and H.H.Yeung are with the Department of Linguistics,
        Simon Fraser University}%
\thanks{$^{4}$A.Basic is with the Mather Institute, 1560 Sherman Avenue, IL, USA}%
\thanks{$^{5}$H.H.Yeung is with the Integrated Neuroscience and Cognition Centre (CNRS and U. Paris Cité), 45 rue des Saints-Pères, Paris, France}%
}

\begin{document}

\maketitle
\thispagestyle{empty}
\pagestyle{empty}

\begin{abstract}

Modern, neural-based ASR and TTS systems for speech communication are now very sophisticated, but they are frequently designed for average users in more idealized scenarios, which can underperform when interacting with users having specialized listening needs, or who speak in non-standard ways. Here we review recent TTS research, and note that industrial, production-level implementations of adaptive TTS systems are scarce. We describe how we implemented changes to TTS systems based on prior research, and then report qualitative results from two case studies, where adapted TTS techniques are applied to real-world HRI settings: First, in an older adult care home, and second, in an airport, where noise levels create comprehension problems. Results suggest promising directions using relatively simple modifications to pitch, duration, and intensity of the robot's speech, and offer a framework for larger, more formal studies of how TTS adaptations can improve HRI.  

\end{abstract}

\section{INTRODUCTION}

Current advances in AI raise many ethical issues, be it the replacement of workers \cite{vorobeva2022thinking, rion2025ai}, environmental issues \cite{green, luccioni2024environmental}, or economic concerns \cite{youvan2025dot, storm2025ai}, so it is critical that future AI developments improve the human experience. A barrier to this goal is that current AI training strategies are centered around the ``average'' user.

\begin{figure}[t]
  \centering
  \includegraphics[width=0.8\linewidth]{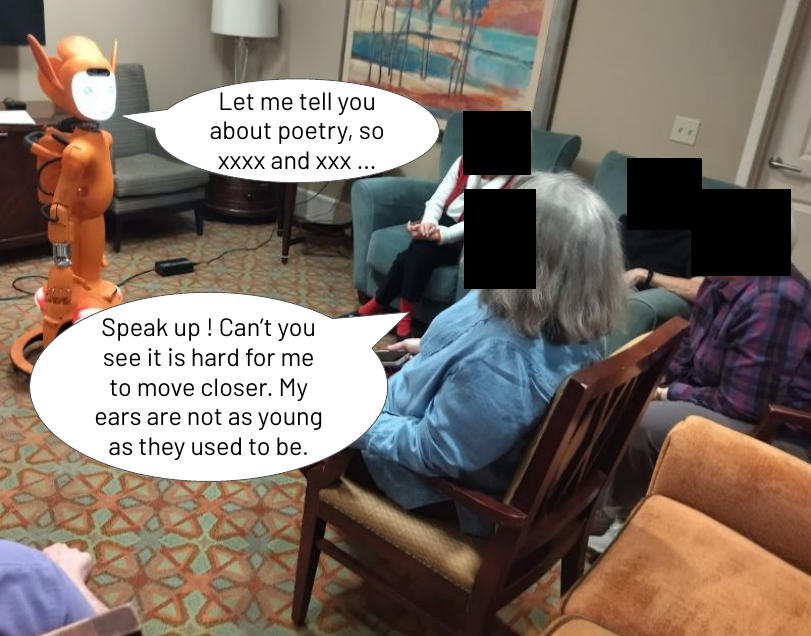}
  \caption{A Miroka robot is speaking too softly for an older adult audience. As the volume does not automatically adjust, the residents must move closer, making the human meet the technology, rather than the technology meeting the human.}
  \label{fig:cover}
  \vspace*{-5mm}
\end{figure}

 The human-robot-interaction (HRI) community must consider adapted systems when interacting with diverse users (Fig. \ref{fig:cover}). In the speech domain, such adaptations may include automatic speech recognition (ASR) and text-to-speech (TTS) systems that consider socio-demographic variables such as gender, ethnicity or dialect \cite{Zellou2024}, and that can adapt to users with special challenges, such as non-native language speakers \cite{kaiser2024artificial} or populations with hearing deficits. Specifically, we focus here on TTS adaptations at several linguistic levels--including acoustic-phonetics, phonology, syntax, and pragmatics--of produced synthetic speech in relation to the perceptual capacities--both normal and adaptive--of human listeners. 

We envision adaptive TTS developing in two stages. First, we must develop new computational methodologies that combine human-centered experimental and generative algorithms to steer speech synthesis in directions that facilitate smooth and inclusive interactions. Second, these adaptations must then be deployed in an existing social robotics platform and tested in semi-experimental real-world situations, with a view not simply to achieve a degree of acceptability using standardized measures, but also to document whether and how truly adaptive robots improve social interactions.

Here, we present two case studies that explore the currently limited work in adapted TTS, then apply a few of these methodologies in a real-word, production-level robotics interaction with users. We describe the initial difficulties in applying out-of-box systems designed for ``default users'', document improvements employed, and report initial feedback on how small adaptive changes have improved the relevant interaction. This work stands as an initial exploration into how implementing even incremental adaptive capabilities can improve interaction for users outside the ``default.'' In short, these studies represent a call to action for \textbf{real-world TTS systems that adapt to humans, rather than expecting humans to adapt to TTS.}

\section{ADAPTIVE TTS: \emph{Where are the gaps?}}

To the best of our knowledge, no commercial TTS systems reliably adapt their output to the listener or listener environment. This, however, should not be a goal: Technology should come further to meet our needs. Although speech technology is regularly referenced as reaching ``human-like'' abilities both in terms of perception (ASR \cite{asr_human_noise, mansfield21_interspeech}) and generation (TTS \cite{lavan2025voiceclones, warning2023deepfakes, srinivasavaradhan25_interspeech}), these systems are frequently assessed by users with ``normal'' speaking and hearing, requiring users outside this group to adapt. For example, both adults and children produce machine-directed speech (e.g. to a smart speaker such as Amazon Alexa) that differs from human-direct speech, with slower speech rate duration and higher pitch \cite{cohn2024technology}. Furthermore, there is a rising concern that the continued use of these types of human adaptations will permanently make their way into speech habits \cite{a0b0acf73ea34a318021d354df2457d4,GUER24}, changing the way we communicate with other humans. Lastly, initial research on human users with reduced comprehension suggests that the modifications to TTS systems required to accommodate these individuals are not harmful to the experience of ``average'' users \cite{preference_age}.

Research in human conversations has demonstrated that effective communication often involves interlocutors adapting to each other, rather than unidirectional adjustments \cite{olmstead2021phonetic, tupper2018game}. Often, humans adapt to create ``deliberately clear speech'' \cite{confluent_talker_and_listener} with a variety of goals including being heard in difficult listening environments \cite{vocal_effort_acoustic_environments, acoustic_phonetic_adverse_listening_conditions}, being listener-specific--as is the case in infant-directed \cite{Smith2024}, hearing-impaired-directed \cite{mommy_speak_clearly}, and machine-directed speech  \cite{prosodic_and_intelligibility}--or deciding in real time when the listener can perceive when they are not being understood \cite{Biro22, Maniwa09, Banse96, Bachorowski99}. These findings underscore the importance of bidirectional adaptation in  human-robot interaction, suggesting that TTS systems should not only be intelligible but also dynamically adapt to users' speech patterns and communicative needs. Although these human modifications are well-studied and something we expect from our conversational counterpart, research on conversational agents, and even more so in social robotics, focuses instead primarily on human-likeness and naturalness, which remain poorly defined \cite{pandey25_ssw, NUSSBAUM2025467, dall14_speechprosody, PERROTIN2025101747}.

While human adaptive speech intends to enhance intelligibility, non-adaptive, or perhaps even worse, maladaptive conversation systems used for robotic interaction, are an important factor in the breakdown of real-world use of these technologies. This is especially evident in populations that fall outside of the ``average'' for listening and conversational abilities, such as children, people with disabilities using adapted communication modes, or older adults. In the domain of text-to-speech synthesis (TTS), listener diversity has mainly been addressed through language selection and building speech architectures that support multi-language datasets \cite{coqui}. Yet, even within a single language, evidence from linguistics and cognitive science research show that human speakers have a large variety of ways to adapt their speech to context in order to facilitate their communication goals with their interlocutor \cite{IVEY71}. 

Because human adaptations are so complex and because TTS engineers design and test for ``default users'' expecting other users to adapt themselves, we observe a general lack of interest, especially in production-level systems, to produce adaptive and inclusive TTS. As tends to be the case in adaptive ASR \cite{access_asr2}, smaller models that are easy to train and require small amounts of data open up possibilities in adaptive TTS, where data for particular groups or languages may be difficult to collect. Matcha-TTS \cite{mehta2024matcha} is one particular TTS that has seen explosive usage in expanding pronunciation editing \cite{shim25_interspeech, li25_ssw, tannander24_interspeech}, use in low resource languages \cite{norway_dialect, lodagala25_interspeech}, and in use-case-specific expressivity \cite{emojivoice}. In one particular application of Matcha, we provided the first TTS tailored to a ``non default'' user group in the form of a second language speaker directed TTS \cite{tuttosi25_ssw}. Moreover, free open-source models remain particularly important to the HRI research community and an important factor in inclusivity as it allows modern TTS systems to be used by those with limited resources and budgets. State-of-the-art TTS systems, while high performing, can cost hundreds of dollars per month even with a small number of users\footnote{https://elevenlabs.io/pricing}.

If TTS research itself is not driving forward the need for adaptive and inclusive systems, this may be because  research studies are often constructed in idealized environments, free of having to truly see the breakdown in communication when applied with real world users. As such, we suggest that HRI studies provide an excellent avenue to pinpoint these breakdowns and explore even simple adaptations that improve equity resulting in smoother human-robot interactions.

\section{CASE STUDIES}
We explore two use cases from social robotics. The first is about human \emph{comprehension challenges}, both in older adults and those living with disabilities, and the second is about environmental challenges in \emph{noisy environments}.

Each use case will be presented as follows: First, we present the initial observed conversational breakdown; next we present our adaptive modification to the system; then we present follow up observations on the improvements, or lack there of, of the implemented system. 

\subsection{Human Comprehension Challenges}

\subsubsection{Initial Breakdown}
Our first experience in communication breakdowns due to human comprehension limits occurred in a care community in the United States. Through initial, informal interactions with users, we saw multiple shortcomings of our conversational system. The ASR, the TTS, and especially turn-taking systems struggled to adapt to ``non default'' cases of older adult users. Often, these users users have slower speech \cite{janse2009processing}, leading to interruptions and misunderstandings. Although these are important future problems to be solved, we will focus our work here on limitations and proposed adaptations for TTS.

In these initial non-structured interactions with senior care residents, we saw almost all residents remarked ``they are just too fast for me," and responding to the robot ``I don't understand what you are saying." Residents reported the voices were both too fast, and too quiet for them to understand, resulting in them needing to move closer to the robot or asking for several repetitions.

\subsubsection{TTS Modifications}
For this first study, our Mirokaï robots\footnote{https://enchanted.tools/} used an in-house TTS based on FastPitch \cite{fastpitch} in both English and French. This small scale TTS allows us to have the system embedded in the robot so that we do not need to rely on fast internet connectivity that is often lacking in both airports, retirement homes, and many other use-case locations for social robots.

To address the problem of speech rate, we applied an overall slowing down to 80\% speed via wizard-of-oz techniques (WOZ). SoX\footnote{https://sourceforge.net/projects/sox/} was used to adjust the speech rate of the wav file before it was played.

To address the problem of volume control we use our \textbf{noisy mode} initially implemented for the \emph{noisy environment} scenarios. This adaptation includes a reduction in speech rate and increase in pitch as volume increases. For a more in depth explanation see section \ref{noisy-adapt}.

\subsubsection{Follow-Up Study 1 - Older Adult Care}

Our sample of residents in a care home in the USA (\textit{N} = 27; \textit{mean age} = 83.5 years) participated in either a group activity of 2-6 people or 1-on-1 interactions (\textit{N} = 8). The 1-on-1 sessions ranged from 8-10 minutes and involved mood checks or discussions of the residents' memories. The group activities lasted 15-20 minutes and included upper-body workouts, name-that-tune games, movie talk, trivia quizzes, riddle games, and art talks.

\begin{figure}[t]
  \centering
  \includegraphics[width=0.7\linewidth]{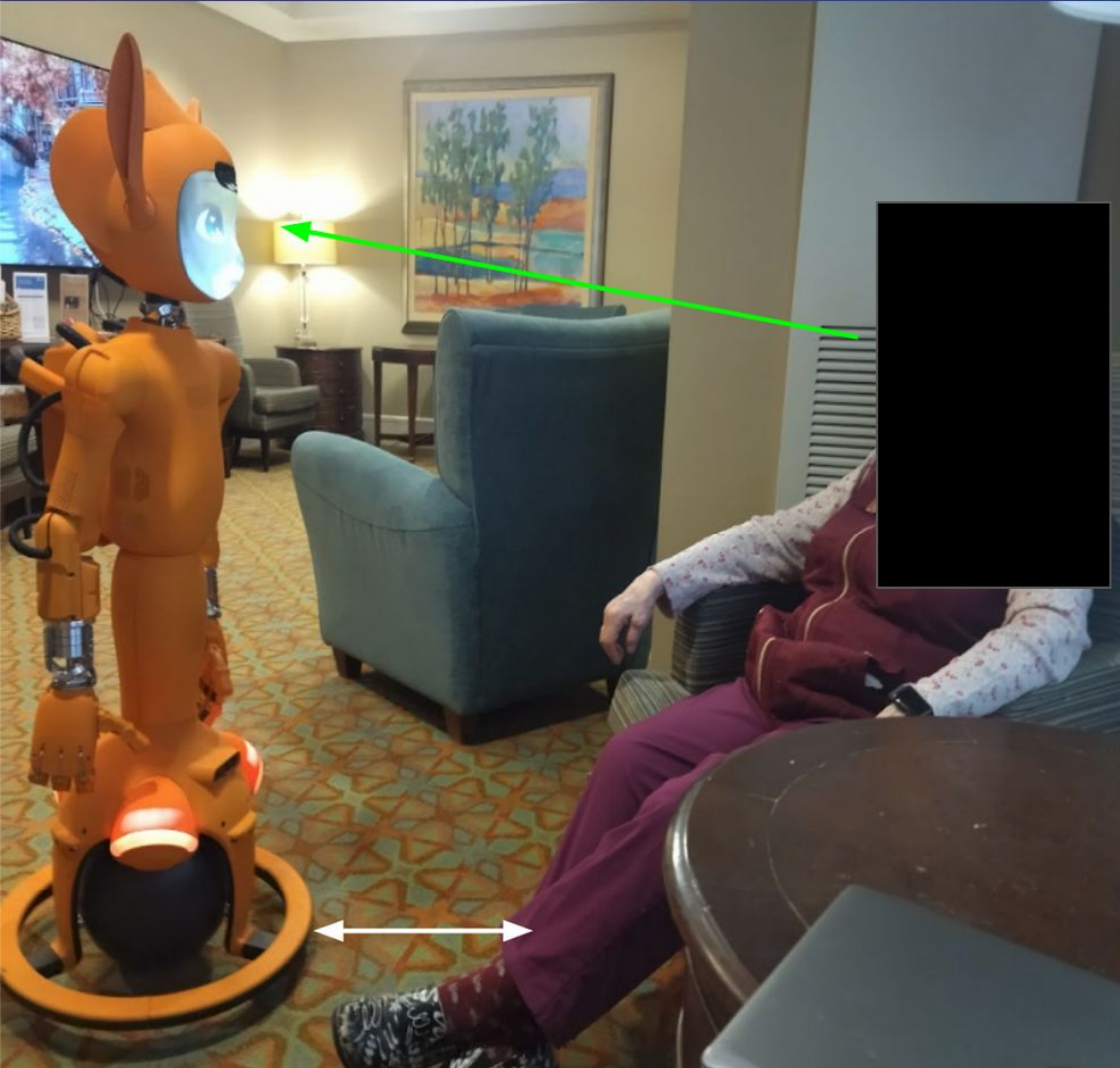}
  \caption{Position of robot and resident in 1-on-1 interaction in the retirement care use case.}
  \label{fig:1on1}
  \vspace*{-2mm}
\end{figure}

\begin{figure}[t]
  \centering
  \includegraphics[width=0.65\linewidth]{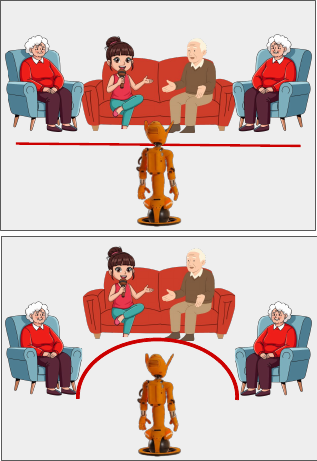}
  \caption{Position of the robot, residents, and facilitator for group interactions in the USA retirement care use case. Top: line facing the robot, Bottom: semi circle around the robot, with the facilitator always centered.}
  \label{fig:group}
  \vspace*{-5mm}
\end{figure}

The set-up for the 1-on-1 interactions can be seen in Fig. \ref{fig:1on1}. Residents were seated (or standing with a walker in one case) and positioned between 0.3-1.2m in front of the robot depending on their comfort. Most participants relaxed into their chairs, but some leaned towards the robot, and the resident's eye level was just below that of the robot's. For the group interactions, the residents were seated around the robot either in a line facing the robot (Fig. \ref{fig:group}, top) or in a semi-circle surrounding the robot (Fig. \ref{fig:group}, bottom). In both cases, the robot was placed 0.5-1.5m in front of the residents.

\emph{i) Objective Responses}
We collected daily 5-point Likert scales. These results are combined between both activities, although most residents participated in both activities. The results can be seen in Table \ref{tab:likert_results_maher}. Despite inconsistencies in the conversational system, we had an overall positive resident response to the robot's presence in the senior care center. The lowest rating was related to the ability of the robot to help the residents connect with other residents. This may come from the fact that there were many 1-on-1 sessions and is more likely an activity design flaw rather than a reflection on the conversation system, as we see the overall enjoyment and comfort remained high.

\begin{table}[h]
\centering
\begin{tabular}{lcc}
\hline
\textbf{Question} & \textbf{Mean} & \textbf{SD} \\
\hline
I enjoyed interacting with Mirokai & 4.22 & 0.70 \\
I felt comfortable around Mirokai & 4.59 & 0.50 \\
I felt safe around Mirokai & 4.74 & 0.45 \\
Mirokai made activities more interesting & 3.85 & 0.99 \\
Mirokai helped me feel connected with others & 3.59 & 1.28 \\
I would like Mirokai to visit again & 4.11 & 0.93 \\
\hline
\end{tabular}
\caption{Resident evaluation for senior care interaction with Miroka (Likert scale 1-5, strongly disagree - strongly agree).}
\label{tab:likert_results_maher}
\vspace*{-5mm}
\end{table}

\emph{ii) Subjective Observations}
Although there were far fewer concerns with the robot being difficult to understand, some residents still found the robot spoke too fast. This shows that even among one user group, there is high inter-individual variability in preference for the speaking rate and there is a need for user-controlled speech. Additionally, residents found that the robot's voice was too high-pitched at times. The robot is designed to be expressive; however, different hearing abilities can change the perception of pitch and sensitivity to pitch \cite{Moore2005, hake2025recordedmusic, souza2002compression, distadio2025doyouhear}. This suggests that user-specific expressivity should also be designed, possibly using a system similar to \cite{emojivoice}. Moreover, the voice was often not loud enough for every listener, and the users did not have their own control to change the volume.

\subsubsection{Further TTS Modifications}
In \cite{tuttosi24_interspeech, tuttosi25_ssw}, we found that for second-language speakers, another group with reduced comprehension ability, slowing down speech rate allowed somewhat improved comprehension, but could also result in negative attitudes and comprehension errors in specific words. By not slowing the entire speech at the same rate and instead slowing only targeted words and sounds, we were able to both improve comprehension for second language speakers and maintain a respectful-sounding voice. Following this work, we used the same targeted slowing down methodology and added additional voice commands to allow users to tell the robot to speak slower, faster, louder or softer. Our in-house embedded TTS was updated to follow the architecture of Matcha-TTS \cite{mehta2024matcha}, allowing both phoneme level duration control and the application of these changes prior to vocoding, resulting in cleaner audio manipulations.

\subsubsection{Follow-Up Study 2 - Adults with disabilities}
Following up on these new changes, we visited a residential care home for adults with disabilities with varying levels of autonomy and communication modes (verbal and non-verbal). This follow up was less structured, with the robot interacting with at least 15 residents in both 1-on-1 and group interactions (5-7 participants). The activities included: storytelling, meditation, and relaxation and breathing exercises.

Although struggles with the ASR and turn taking remained, a WOZ ``hold'' feature blocking the ASR did improve the situation. With the implementation of voice-based controls on speech features and targeted slowing, we saw no issues with the residents not understanding what the robot was saying (although it was not the same target population as in the previous follow up). The residents did, however, continue to struggle with the higher pitched tones in the robot, which may be exacerbated through the use of the ``noisy-mode'' initially designed to improve listening ease. 

These changes were not particularly complex to implement and came from studies designed to understand how to conform TTS systems to the user, rather than the reverse. Here, we see evidence that designing for specific user groups and allowing users personal control over their robot companions can greatly improve human robot interactions, especially in ``non default'' user groups.

\subsection{Noisy Environments}

\subsubsection{Initial Breakdown}
\begin{figure}[t]
  \centering
  \includegraphics[width=0.6\linewidth]{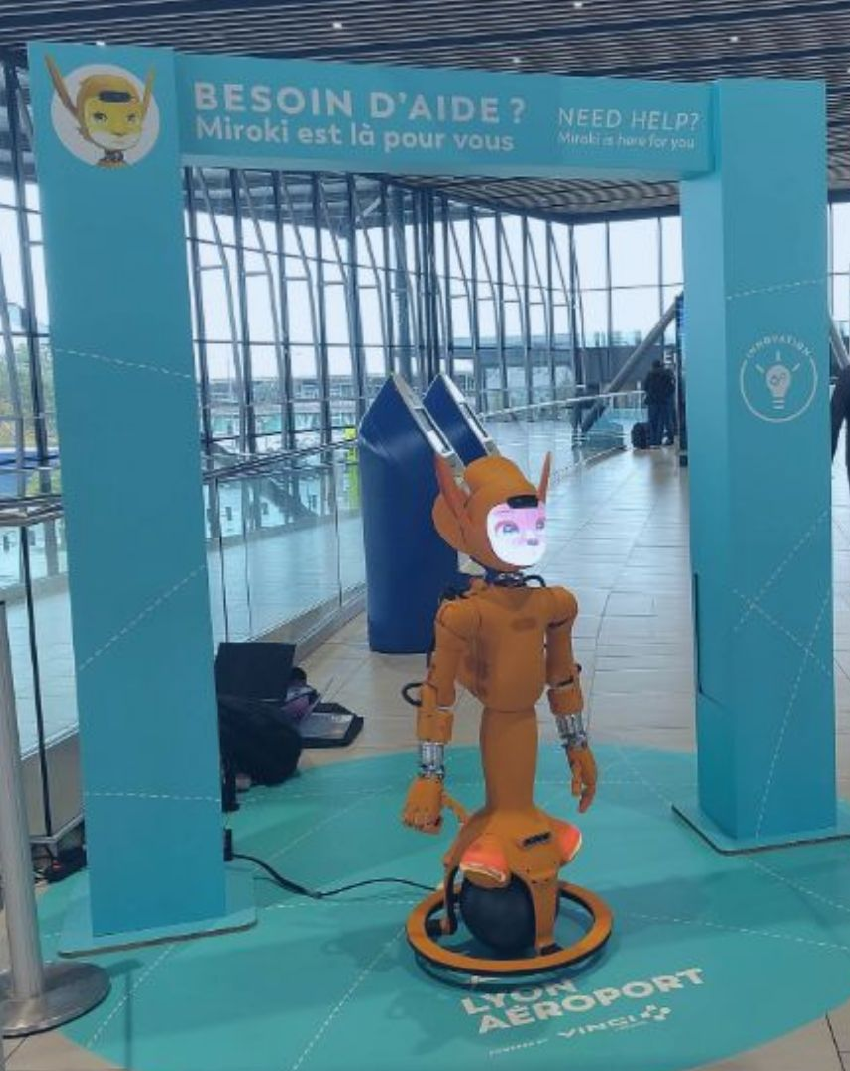}
  \caption{Setup of the robot greeting booth at the Lyon airport.}
  \label{fig:lyon}
  \vspace*{-6mm}
\end{figure}

Our second use case took place in France at the Lyon airport. There were \textit{N}=263 observed interactions and N=138 users provided feedback. The breakdown of passenger type can be seen in Table \ref{tab:passenger_profiles}. The robot was stationed in the airport with a large booth so it could be easily spotted (Fig. \ref{fig:lyon}). The robot provided fun greetings, directions within the airport, and flight information to passengers. The users stood 0.5-1.5m from the robot looking down to the robot's eye level, and the robot was able to orientate its head upwards to meet the passenger's gaze.

\begin{table}[h]
\centering
\begin{tabular}{lcc}
\hline
Passenger Type & Percentage (\%) \\
\hline
Family with children &  40.6 \\
Family or group & 25.0 \\
Person alone & 19.0 \\
Couple &  9.1 \\
Airport staff & 4.5 \\
Senior & 0.3 \\
Person with reduced mobility & 0.3 \\
Not specified & 0.7 \\
\hline
\end{tabular}
\caption{Profile of individuals or groups interacting with the Mirokaï robot at Lyon Airport.}
\label{tab:passenger_profiles}
\vspace*{-6mm}
\end{table}

\emph{i) Objective Responses}
The robot was well received by both staff and passengers. Children were particularly interested in the robot. Users were asked to fill out a questionnaire with 5 point Likert scores (Table \ref{tab:lyon_airport_likert}) and free-form answers, giving feedback on their experience with the robot. In the responses, there were 82 out of 138 mentions referring to the robot as ``cute,'' ``funny,'' ``adorable,'' or ``an attraction.'' There were 87 mentions of the robot's usefulness. Of these, 47 mentioned help with airport services and 28 mentioned help with navigation or path-finding. We noted that for flight-related questions, however, attitudes were more negative, with 65\% of users noting that the robot was bad, or very bad. Lastly, we observed the facial expressions of ``non-interactors,'' i.e. passengers who did not ask the robot a question,  when viewing the robot. Of these ``non-interactors'', 75\% smiled or laughed when they saw the robot. We saw that 8\% of respondents (N = 120) said the robot's response was not understandable and an additional 7.5\% said that the interaction was not easy, but 14.4\% of verbally-interviewed passengers mentioned a negative experience with the robot, with the vast majority citing comprehension problems.

\begin{table}[h]
\centering
\begin{tabular}{lccc}
\hline
Measure & Mean & N \\
\hline
Ease of interaction & 3.77 & 138 \\
Relevance of answers & 3.74 & 119 \\
Satisfaction with interaction & 4.07 & 27 \\
Perceived usefulness of response & 4.00 & 120 \\
\hline
\end{tabular}
\caption{Passenger evaluation from the Lyon Airport interaction with Miroka (Likert scale 1-5, scale depends on question from not at all to very). Standard deviations are not supplied as individual counts are not available.}
\label{tab:lyon_airport_likert}
\vspace*{-8mm}
\end{table}

\emph{ii) Subjective Observations}
There were frequent issues with the robot both not being understood by users, more so than are reflected in the number of objective responses noting non-understanding. As with older adult users, this results in turn-taking issues. For the TTS, the primary pain-point was volume; despite the robot being set to max volume, it was often difficult to hear over the background noise of such a busy environment. Moreover, the high volume saturates the speakers, leading to reduced audio quality.

Despite these difficulties, the response was overwhelmingly positive, especially for users who did not ask high stakes flight-related questions. It is possible that in an interaction that is short-lived and low risk such as these (flight questions excluded), compared to the long-term conversations of the older adult users, the voice quality is less important to users as the novelty effect remains high and they are excited just to be speaking with a robot.

\subsubsection{TTS Adaptations}\label{noisy-adapt}
We encountered several problems in which users struggled to hear the robot despite using highest volume level. Users needed to ask the robot to repeat itself several times in some cases, which often led to trip-ups and disengagement from the user. As such, we saw a strong need for a voice that is more audible in noisy and crowded environments. 

There has been a small amount of research on environmental adaptations for robot voices \cite{volume_adaptation_telepresence_system, renclarity, readroom, HRI2026noisy}. We follow the suggestions in our work \cite{readroom} and \cite{hear_speak} by creating a \textbf{noisy mode} that decreases the speech rate to 75\% and adds a small linear increase in pitch with volume, in the range of 80\% of the maximum volume. The increase is a simple linear formula where the pitch increase (transformed using SoX) is linear from 75 to 90 cents as the volume increases from 75-100\%, represented as: 

$p(V) = P_{\min} + (V - V_{\text{limit}})\frac{P_{\max} - P_{\min}}{100 - V_{\text{limit}}}$
where $P_{\min}$ = 75 cents and $P_{\max}$ = 90 cents.

\subsubsection{Noisy Environment - Follow up}
An official follow-up study to test noisy mode remains future work; however, in initial demonstrations the noisy mode appears much easier to understand in noise without saturating the speakers, and we have received far fewer complaints of not being able to hear the robot or breakdowns in conversation with moving closer, or leaning into the robot and demands for repetition. Moreover, we did not receive the same feedback that the robot was using an uncomfortably high pitch that we saw for the user comprehension use cases. This suggests that pitch augmentation due to volume difficulties is specifically suitable for environmental challenges rather than user comprehension volume-based challenges.

\section{LIMITATIONS and FUTURE ADAPTATIONS}
Although we were able to improve the interlocutor's listening experience in both sets of case studies, our work here is limited to TTS. Yet, many more improvements need to be made to the conversational system to improve the ASR and turn-taking systems before interactions can be truly fluid.

In future work, we plan to implement work similar to our study \cite{shim25_interspeech} to improve residents' complaints with poorly articulated punctuation. Although we applied the work in \cite{tuttosi25_ssw} to address the needs of older adult listeners and those with different comprehension abilities, we recognize that hearing versus comprehension challenges are distinct (e.g., for older adult versus L2 audiences), which suggests that more targeted research is needed. While the current approach has, until now, concentrated on adaptations aimed at L2 speakers, more specific manipulations that can improve voice interactions for each specific user group are likely to exist.

\section{CONCLUSIONS}
There is a general lack of inclusive TTS systems, as they are often designed only with the ``default'' user in mind. We propose, however, that HRI is an ideal venue to explore the need for adaptive TTS systems, as social robotics are working face to face with users and we see, in real time, the breakdown in an interaction that can be caused by a poorly adapted TTS system. Here, we presented an industry perspective of discovering a problem with deploying out-of-box TTS for ``non-default'' users, specifically older adults and in a noisy environment, and deploying TTS from the research community designed specifically for a user group or an environment. We saw, through progressive application of targeted speech rate, volume, and pitch modifications, that we were able to resolve many listening issues for users with comprehension difficulties. We call on the speech and HRI communities to take an interest in expanding systems that are focused on the needs of real user groups who may not have the same needs as the average, because until this is done, social robotics will remain inaccessible to some of the communities who may benefit the most from their use.

\bibliographystyle{IEEEtran}
\bibliography{references}

\end{document}